# Bandwidth tunable optical filter based on a tri-mode high-contrast grating

Hsin-Yu Yao[1], Yi-Chen Wang[1], Tsun-Hsu Chang[1,*], and Tsing-Hua Her[2,*]

[1] Department of Physics, National Tsing Hua University, Hsinchu, 30013, Taiwan

[2] Department of Physics and Optical Science, The University of North Caroline at Charlotte, Charlotte, NC 28223, USA

**Abstract**

We propose a monolithic optical bandpass filter with a transmission bandwidth tunable by the incident angle in a tri-mode high-contrast grating (HCG). We attribute this extraordinary phenomenon to the destructive interference between the $TM_0$ and $TM_2$ modes and the Fabry-Pérot resonance of the $TM_1$ mode, which can only be excited at the oblique incidence. The transmission bandwidth can be tuned from zero at normal incidence and increases quadratically with incident angle, while the center wavelength reds shifts slightly. The corresponding Q factor can be continuously tuned from $10^6$ to $10^2$ as the incident angle increases from nearly $0°$ to $10°$.

It is well known that an optical bandpass filter with a tunable center wavelength can be easily realized by tilting a compact thin-film Fabry-Perot filter [1] or a ¼-wave Bragg stack with a defect layer [2]. A compact optical bandpass filter with a tunable transmission bandwidth, on the other hand, remains challenging. Nowadays, such a device has been demonstrated by a combination of diffractive grating and a shaped reflector [3], a combination of long-pass and short-pass filters [1], Mach-Zander interferometers [4] or cascaded microring resonators [5, 6]. These devices are conceptually simple but suffers either large footprint and/or slow tuning speed. Various applications have been proposed, including channel subset selection and dynamic bandwidth allocation in wavelength-division multiplexing (WDM) for optimal spectral efficiency [3-5], reconfigurable filters for grid-less networking [7], optical performance monitoring [8], and adaptive filtering for signal processing. In this work, we propose a compact monolithic optical bandpass filter with a tunable bandwidth simply by tilting a high-contrast grating (HCG). Previously HCG has been explored as a two-mode Fabry-Pérot interferometer, either at normal or oblique incidence [9, 10], where high reflectivity (> 99%) over a wide wavelength range (fractional bandwidth > 30%) [11, 12] or sharp Fano resonance with Q > $10^6$ [12, 13] were demonstrated. Our tunable bandwidth filter (TBWF) is based on a tri-mode HCG where its transmission bandwidth can be tuned quadratically from zero at normal incidence to a fraction of grating period at an increasing incident angle. We attribute this extraordinary phenomenon to the destructive interference between the $TM_0$ and $TM_2$ modes and the Fabry-Pérot resonance of the $TM_1$ mode, which can only excited at the oblique incidence. The principle and performance of TBWF will be presented.

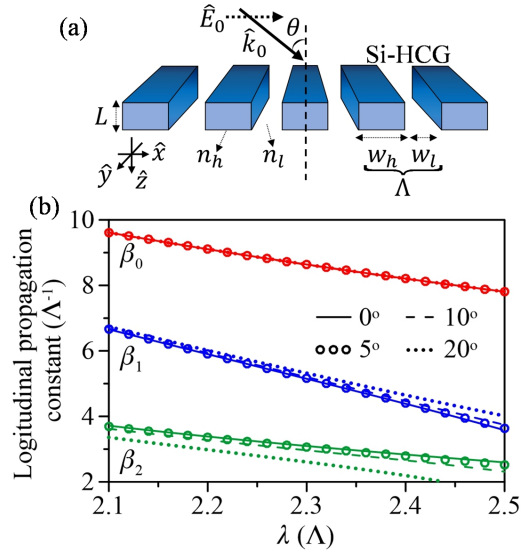

Fig. 1. (a) Schematics of a Si-air HCG immersed in air. A TM-polarized plane wave with the electric field perpendicular to the grating bar ($\hat{E}_0 \parallel \hat{x}$) is incident at the angle of $\theta$. (b)

Dispersion relation of the three propagating modes for the Si-air HCG with duty cycle (DC) of 0.77 over the specified spectral range.

Consider a generic HCG shown in Fig. 1(a), consisting of periodic low-index groove ($n_l$ and $w_l$) and high-index ridge (index $n_h$ and width $w_h$) with a period of $\Lambda = w_h + w_l$ and a length of $L$. In this work, we illustrate this principle using a Si-air HCG ($n_h = 3.476$ and $n_l = 1$) immersed in air, excited by a TM-polarized wave ($\hat{E}_0 \parallel \hat{x}$) at the incident angle $\theta$. Due to the oblique incidence, the waveguide array modes (WAMs) have transverse propagation component ($\parallel \hat{x}$) satisfying pseudo-periodic boundary condition that is characterized by a Bloch wavenumber $k_{x0} = k_0 \sin\theta$ with $k_0 = \omega/c$, the free-space wavevector [14]. The magnetic field within the $m$-th unit cell of the grating ($0 \le x < \Lambda$) can be written as [14, 15]

$$H_y(0 \le x < \Lambda, z) = \sum_j h_{y,j}(x) e^{ik_{x0}m\Lambda} \left( f_j e^{i\beta_j z} + b_j e^{-i\beta_j z} \right), \tag{1}$$

where $f_j$ and $b_j$ denote respectively the forward-wave and backward-wave coefficients of the $j$-th WAM along the $z$ direction with a propagating constant $\beta_j$. $h_{y,j}(x)$ is a periodic function that carries the translational symmetry of the lattice and can be expressed by

$$\begin{cases} h_{y,j}(0 \le x < w_l) = A_j \cos\left[k_{l,j}(x - w_l/2)\right] \\ \qquad\qquad + B_j \sin\left[k_{l,j}(x - w_l/2)\right] \\ h_{y,j}(w_l \le x < \Lambda) = C_j \cos\left\{k_{h,j}\left[x - (w_l + \Lambda)/2\right]\right\} \\ \qquad\qquad + D_j \sin\left\{k_{h,j}\left[x - (w_l + \Lambda)/2\right]\right\} \end{cases}, \tag{2}$$

where $k_{l,j} = \sqrt{n_l^2 k_0^2 - \beta_j^2}$ and $k_{h,j} = \sqrt{n_h^2 k_0^2 - \beta_j^2}$. By matching the lateral boundary conditions [14], the ratios of $B_j$, $C_j$, and $D_j$ to $A_j$ can be obtained (see Eqs. S3-S5 of Supplement 1), and the propagating constants $\beta_j$ can be shown to satisfy the following dispersion relation [15]:

$$\begin{aligned}&\cos(k_{l,j}w_l)\cos(k_{h,j}w_h) \\ &-\frac{1}{2}\left(\frac{n_h^2 k_{l,j}}{n_l^2 k_{h,j}} + \frac{n_l^2 k_{h,j}}{n_h^2 k_{l,j}}\right)\sin(k_{l,j}w_l)\sin(k_{h,j}w_h) = \cos(k_{x0}\Lambda)\end{aligned}. \tag{3}$$

Figure 1(b) displays the dispersions of the three propagating WAMs in the spectrum of interest, including the two even modes ($TM_0$ and $TM_2$), and one odd mode ($TM_1$), for incident angle between 0º – 20º. Note that for surface-normal incidence, $TM_1$ mode cannot be excited due to symmetry mismatching. Figure 1(b) indicates that the sensitivity of the modal dispersion on the incidence angle increases with the mode order: $\beta_0$ stays nearly independent of $\theta$ while $\beta_2$ displays more pronounced change at increasing the incident angle. The reflectance of the HCG is solved using multimode Airy analysis with 51 (3 propagating and the rest evanescent) modes for better accuracy [16]. Alternatively, it can also be calculated using transfer matrix method [9]. All results shown here were numerically validated by the Rigorous coupled-wave analysis (RCWA) and Ansys HFSS (High-frequency Structure Simulator, ANSYS, Inc.).

The reflectance of the HCG depends on the length (in the unit of the grating pitch, $L/\Lambda$) and the duty cycle (defined as $DC \equiv w_h/\Lambda$) of the grating, the light wavelength (in the unit of the grating pitch, $\lambda/\Lambda$), and the incident angles $\theta$. Figure 2(a) and 2(b) show the reflectance spectrum for various $DC$ (at a fixed $L = 0.61\Lambda$) and $L$ (at a fixed $DC = 0.77$) for incidence angle $\theta$ below 15°. For normal incidence ($\theta = 0°$), broadband reflectance appears from $\lambda = 2.1\Lambda$ to $2.5\Lambda$, corresponding to a fractional bandwidth larger than 15%. According to Karagodsky et al., this broadband reflection is understood by the destructive interference of $TM_0$ and $TM_2$ at the exit interface [12]. As $\theta > 0°$, a high-transmission channel opens, whose bandwidth is sensitively dependent on incident angles. The larger the $\theta$, the broader the transmission bandwidth. Reflectance spectrum at incident angles larger than 15° exhibits even larger opening but less contrast with a deformed shape (see Fig. S1). Another interesting observation is that the transmittance maximum red shifts with either increasing $L$ or increasing $DC$. This suggests the transmission peak is in a form of optical resonance with a well-defined phase. As the $DC$ or $L$ increases, the resonance wavelength $\lambda_r$ needs to increase (or red shifts) in order to maintain a constant modal propagating phase ($\beta_j L$), where a larger $DC$ results in higher effective index of each WAM (i.e., $n_{\text{eff},j} \equiv \beta_j/k_0$). This conjecture is further supported by Fig. 2(c) which shows 2D reflectance map as a function of the grating length and the duty cycle at a light wavelength $\lambda = 2.3\Lambda$ for various incident angles $\theta$. Here again a transmission peak appears and widens for increasing incidence angles. The $DC$ and the corresponding $L$ for which the high transmission occurs appears anti-correlated, i.e., $DC \cdot L \sim const.$, suggesting a constant modal propagating phase $\beta_j L$ for the observed high transmission. This again indicates that this high transmission is an optical resonance in nature.

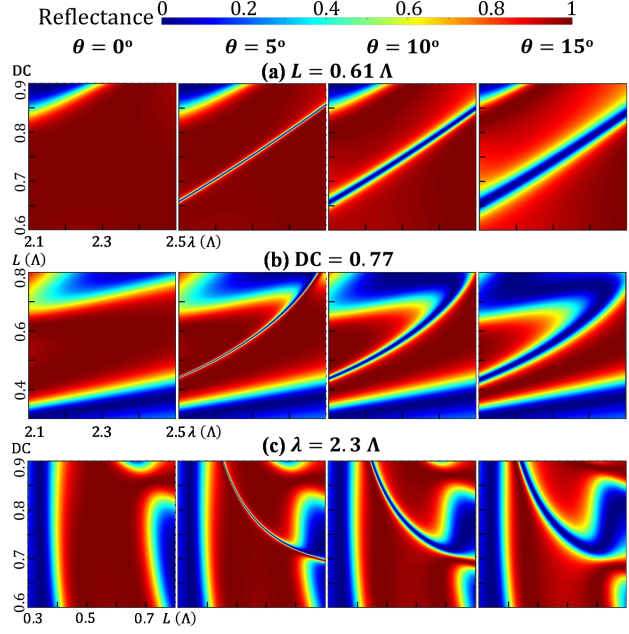

Fig. 2. (a) 2D reflectance maps as a function of wavelength ($\lambda$) and duty cycle (DC). $L$ is fixed at $0.61\Lambda$. (b) 2D reflectance maps as a function of wavelength ($\lambda$) and grating length ($L$). DC is fixed at $0.77$. (c) 2D reflectance maps at $\lambda = 2.3\Lambda$ as a function of grating length ($L$) and duty cycle (DC). Subplots from left to right in every row are the results with $\theta$ gradually varying from $0°$ to $15°$.

Perhaps the most direct proof of the resonance nature of this sharp passband is to examine the self-sustainability of the mode associated with this high transmission. Karagodsky *et al.* has demonstrated that high-$Q$ resonance in HCGs occurs when the assembly of WAMs, so-called the supermode, becomes self-sustainable [13], *i.e.*, $\det[\mathbf{I} - (\mathbf{p}\,\mathbf{r}')^2] = 0$, where $\mathbf{I}$ is the identity matrix, $\mathbf{p}$ is the diagonal propagation matrix with elements $p_{jj} = \exp(i\beta_j L)$, and $\mathbf{r}'$ is the internal reflection matrix describing the reflection of each of WAMs into themselves and their coupling into other modes [9, 16]. To see this, a HCG with $DC = 0.77$ and $L = 0.61\Lambda$ is selected for analyses. Figure 3(a) plots the corresponding 2D reflectance map as a function of $\lambda$ and $\theta$. It shows the transmission bandwidth increases with the incident angle, while the resonant frequency red shifts slightly. Figure 3(b) plots the corresponding contour of $\det[\mathbf{I} - (\mathbf{p}\,\mathbf{r}')^2]$ for the reduced $3 \times 3$ matrix, where the white dotted line highlights the minimum of $\det[\mathbf{I} - (\mathbf{p}\,\mathbf{r}')^2]$. Since the high transmission band follows the minimum of $\det[\mathbf{I} - (\mathbf{p}\,\mathbf{r}')^2]$, Figs. 3 (a) and 3(b) clearly demonstrates the observed high transmission is a Fabry-Pérot (FP) resonance of the tri-mode HCG.

For the passband in Fig. 3(a), Fig. 3(c) plots its full-width-at-maximum (FWHM) $\Delta\lambda_{3dB}$ and the shift of the resonant wavelength $\Delta\lambda_r$, defined by the resonant wavelength with respect to that at the limit of normal incidence, as a function of $\theta$. The transmission bandwidth increases super-linearly from zero at normal incidence to $3.6\times10^{-2}\Lambda$ at $\theta=10°$, while the resonant wavelength $\lambda_r$ red shifts at a slower rate. The red shift can be understood as follows: in order to maintain the resonance during the angle tuning, $\beta_1 L$ needs to stay at a particular constant value. With increasing the incident angle, Fig. 1(b) indicates the modal dispersion curve red shifts, meaning the same $\beta_1$ requires a longer wavelength. Fig. 3(c) also displays the $Q$ factor ($=\lambda_r/\Delta\lambda_{3dB}$), showing $Q$-factor can be continuously tuned from $10^6$ at $0.1°$ to $10^2$ at $10°$. Fig. 3(d) shows the overall electric-field intensity of the resonance supermode for an incident angle of 1°, normalized to the incident intensity. It is a first-order resonance, since the overall intensity exhibits one peak along the $z$ axis. An energy buildup of more than $10^4$ times is observed here.

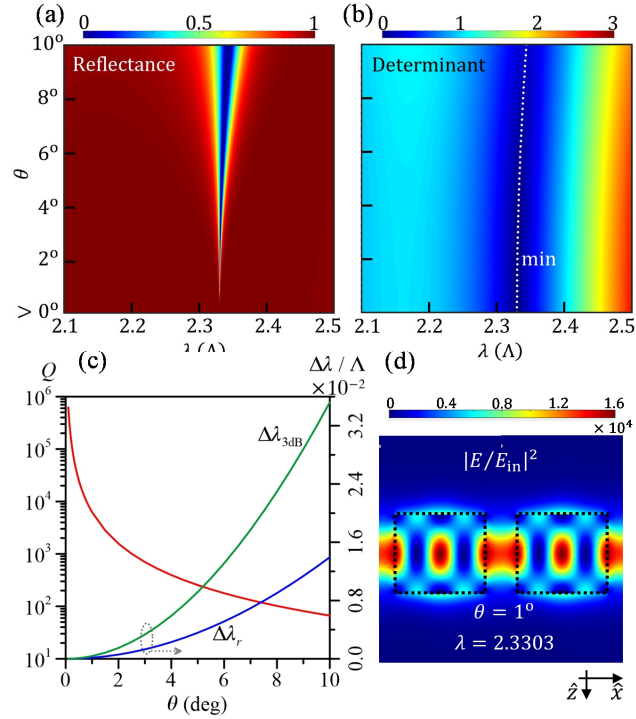

Fig. 3. (a) 2D reflectance map as a function of wavelength ($\lambda$) and incident angle ($\theta$). $L$ and DC are $0.61\Lambda$ and $0.77$, respectively. (b) 2D determinant map ($|\det[\mathbf{I}-(\mathbf{p}\,\mathbf{r}')^2]|$) in the same parameter space as (a). (c) $Q$ factor (left-handed axis), resonant wavelength shift ($\Delta\lambda_r$, right-handed axis), and 3dB bandwidth ($\Delta\lambda_{3dB}$, right-handed axis) versus incident angle $\theta$. (d) Normalized field intensity at resonance ($\lambda=\lambda_r=2.3303\Lambda$) when $\theta=1°$.

To examine the nature of this resonance, we denote the eigenvalues of single-trip transfer function $\mathbf{p}\,\mathbf{r}'$ as $\gamma \equiv |\gamma|\exp(\angle\gamma)$. The supermode with the largest $|\gamma|$ will have the strongest feedback to build up energy in the resonator, which will dominate its response [13]. Figures 4(a)-(c) plot respectively the spectra of $|\gamma|$, $\angle\gamma$, and $|\gamma/(1-\gamma^2)|$ of the dominating supermode in the vicinity of resonance as a function of $\theta$. The last quantity is proportional to the Q factor of HCG $Q_{HCG} = \max\left(2\pi n_{gr}(L/\lambda)|\gamma/1-\gamma^2|\right)$ and is dubbed as $Q'_{HCG}$ in this report [13]. While $|\gamma|$ displays a weak variation, $\angle\gamma$ shows a strong wavelength dependence where the black dotted line indicates the contour of $\angle\gamma = \pi$, which matches the minimum determinant contour in Fig. 3(b). Along this contour the supermode interferes with itself constructively after a round trip and exhibits large Q value (Fig. 4 (c)). The reduction of Q factor and hence the broadening of the passband away from normal incidence is due to a decreasing $|\gamma|$ at an increasing incident angle. Modal analysis shows this supermode is primarily composed of the $TM_1$ mode. To see this, Figure 4 (d)-(e) plot the amplitude, phase, and $Q'_{HCG}$ factor for the single-trip transfer function $p_1 r'_{11}$ where $r'_{11}$ is the field reflection coefficient of $TM_1$ mode at the interfaces. As shown, the contour of $\beta_1 L + \angle r'_{11} = \pi$ matches that of $\angle\gamma = \pi$, and the 2D maps of $|r'_{11}|$ and $Q'_{HCG}$ display very similar trends as those of the supermode, especially for $\theta \leq 5°$.

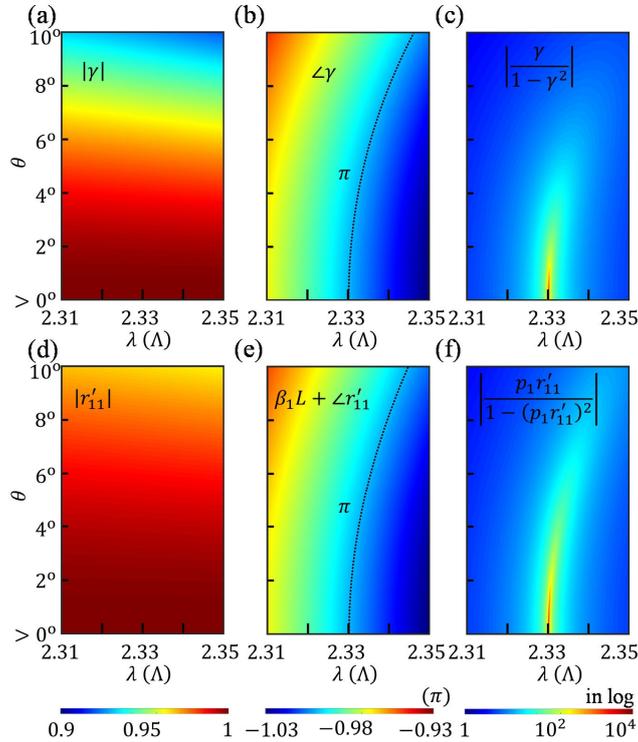

Fig. 4. (a) and (b) respectively plot $|\gamma|$ and $\angle\gamma$ of supermode as a function of $\lambda$ (only around $\lambda_r$) and $\theta$. (c) plots $|\gamma/(1-\gamma^2)|$ calculated according to (a) and (b), which should be proportional to the $Q$ factor. (e)-(f) respectively illustrates the counterparts of (a)-(c) for the single-pass transfer function of the TM$_1$ mode ($p_1 r'_{11} = |r'_{11}| \exp[i(\beta_1 L + \angle r'_{11})]$).

The appearance of this passband can be understood as follows. For plane-wave excitation at the normal incidence, only TM$_0$ and TM$_2$ modes are excited, which can exhibit a broadband reflection when their single-trip phase lapses are off by $\pi$, leading to a destructive interference in the output [12]. At oblique incidence, an additional TM$_1$ mode is excited. For small incident angles, the destructive interference between TM$_0$ and TM$_2$ still holds, contributing to zero output. The TM$_1$ mode, on the other hand, possesses very large reflection coefficient $|r'_{11}|$ (Fig. 4(d)) and hence exhibits a very high-$Q$ Fabry-Pérot resonance. At large incident angle, the reflectivity of the TM$_1$ mode drops and the destructive interference between the TM$_0$ and TM$_2$ is compromised, leading to a smaller energy buildup and broadband transmission.

There is a remarkable difference in the resonance characteristics between the bi-mode HCG reported earlier (Fig. 37 in [10]) and tri-mode HCG reported here. The former is a Fano resonance in nature, resulting from the strong coupling between two propagating modes, excited by normally or obliquely incident light [9, 10, 12]. Each of these two modes is on resonance (even-$\pi$ round-trip phase) and the difference between their single-trip phase is also even $\pi$ [12], leading to a theoretical $Q$ factor as high as $10^9$ [13]. Due to such tight constraints, Fano resonance occur only at discrete locations in the configuration space, where tuning of its center wavelength requires modulation of physical attributes such as refractive index, duty cycle, and the grating length, etc. [16]. The latter is a Fabry-Pérot resonance in nature, resulting from the cancelation of two modes which gives way for the remaining mode with strong reflective feedback to dominate its output. The theoretical $Q$ factor can be very high at very small tilt angle; however, in practice, it will be limited by the device quality and the ability to control and adjust the tilt angle.

To emphasize the possible applications, the reflectance of the HCG of Fig. 3 (a) versus $\lambda$ at seven representative $\theta$ are plotted in Fig. 5(a), showing its potential as a tunable bandwidth filter for a broadband source. In the current design, the center wavelength of the passband also red shifts as the bandwidth broadened, although at a slower rate (See Fig. 3 (c)). As the red shift originates from the TM$_1$ modal dispersion (Fig. 1 (b)), it is possible to engineer the HCG to minimize the red shift. For a narrowband source, Fig. 5 (b) plots the reflectance versus $\theta$ for six wavelengths, showing its potential as a variable neutral density filter, whose OD (optical density) value can be continuously tuned by tilting. At $\lambda = 2.333\Lambda$, the reflectance attenuation can be tuned by varying $\theta$ from $2° - 4.5°$. At $\lambda = 2.343\Lambda$, the theoretical attenuation can be as high as $-80$ dB. HCG is much more lightweight and its tuning mechanism is simpler as compared to [1, 3-6]. By integrating with micro-electro-mechanical systems (MEMS), more compact bandpass filter with faster tuning speed can be realized, which enables a direct integration with a cavity for laser's intensity or frequency modulation.

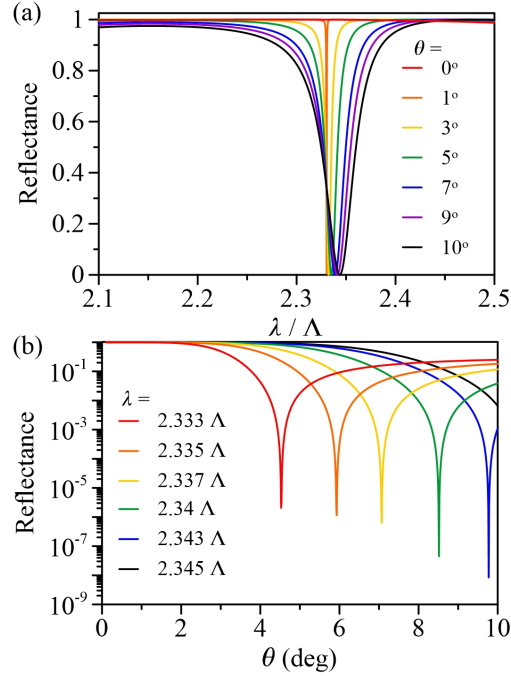

Fig. 5. (a) Reflectance versus wavelength at seven incident angles. (b) Reflectance versus incident angle for six wavelengths.

In summary, we theoretically demonstrate a tri-mode HCG bandpass filter with a transmission bandwidth tunable by incident angle. The proposed HCG metasurface is totally reflective at normal incidence, while it becomes transmissive at oblique incidence. Unlike Fano resonance in a bi-mode HCG, such a phenomenon can be observed over a wide parameter space. For the current design, the center wavelength of the passband red shifts with tilting at a rate that is less than half of the bandwidth tuning. The $Q$ factor continuously varies from $10^6$ at nearly normal incidence to $10^2$ at $10°$. We attribute this extraordinary phenomenon to the destructive interference between the $TM_0$ and $TM_2$ modes and the Fabry-Pérot resonance of the $TM_1$ mode, which can only be excited at the oblique incidence. Our filter is very compact which shows promising applications in agile control and modulation of light bandwidth and intensity.

**Funding sources.** Ministry of Science and Technology, Taiwan under the contract no. MOST 110-2112-M-007-013-MY3 and MOST 110-2811-M-007-552-MY3.

**Discloruses.** The authors declare no conflicts of interest.

See Supplement 1 for supporting content.

# Bandwidth tunable optical filter based on a tri-mode high-contrast grating: supplemental document

## 1. Waveguide-array mode of a HCG excited by oblique-incidence light

Recall that the lateral profile of the magnetic field of $j$-th waveguide-array mode (Eq. (1) of the main text) is

$$\begin{cases} h_{y,j}(0 \leq x < w_l) = A_j \cos[k_{l,j}(x - w_l/2)] + B_j \sin[k_{l,j}(x - w_l/2)] \\ h_{y,j}(w_l \leq x < \Lambda) = C_j \cos\{k_{h,j}[x - (w_l + \Lambda)/2]\} + D_j \sin\{k_{h,j}[x - (w_l + \Lambda)/2]\} \end{cases},$$

The later profiles of the electric-field components $E_x(x,z)$ and $E_z(x,z)$ are denoted as $e_x(x)$ and $e_z(x,z)$, respectively. According to Maxwell's equations, they can be written as

$$\begin{cases} e_{x,j}(0 \leq x < w_l) = \dfrac{\beta_j}{n_l^2 k_0} \sqrt{\dfrac{\mu_0}{\varepsilon_0}} h_{y,j}(0 \leq x < w_l) \\ e_{x,j}(w_l \leq x < \Lambda) = \dfrac{\beta_j}{n_h^2 k_0} \sqrt{\dfrac{\mu_0}{\varepsilon_0}} h_{y,j}(w_l \leq x < \Lambda) \end{cases}, \quad (S1)$$

and

$$\begin{cases} e_{z,j}(0 \leq x < w_l) = \dfrac{k_{l,j} \Lambda}{n_l^2} \sqrt{\dfrac{\mu_0}{\varepsilon_0}} \left\{-A_j \sin\left[k_{l,j}\left(x - \dfrac{w_l}{2}\right)\right] + B_j \cos\left[k_{l,j}\left(x - \dfrac{w_l}{2}\right)\right]\right\} \\ e_{z,j}(w_l \leq x < \Lambda) = \dfrac{k_{h,j} \Lambda}{n_h^2} \sqrt{\dfrac{\mu_0}{\varepsilon_0}} \left\{-C_j \sin\left[k_{h,j}\left(x - \dfrac{w_l + \Lambda}{2}\right)\right] + D_j \cos\left[k_{h,j}\left(x - \dfrac{w_l + \Lambda}{2}\right)\right]\right\} \end{cases}. \quad (S2)$$

The ratios of $B_j$, $C_j$, and $D_j$ to $A_j$ can be solved by matching the lateral boundary conditions and the pseudo periodic condition, including 1) $h_{y,j}(x = w_l)$ and $e_{z,j}(x = w_l)$ are continuous and 2) $h_{y,j}(x = 0) \times \exp(ik_{x0}\Lambda) = h_{y,j}(x = \Lambda)$ and $e_{z,j}(x = 0) \times \exp(ik_{x0}\Lambda) = e_{z,j}(x = \Lambda)$. Those yield

$$\dfrac{B_j}{A_j} = \dfrac{k_{h,j} n_l^2 \cos\phi_l \left[\cos(\phi_0/2)\exp(i\phi_0/2) - \cos^2\phi_h\right] + \dfrac{1}{2} k_{l,j} n_h^2 \sin\phi_l \sin(2\phi_h)}{k_{h,j} n_l^2 \sin\phi_l \left[i\sin(\phi_0/2)\exp(i\phi_0/2) + \cos^2\phi_h\right] + \dfrac{1}{2} k_{l,j} n_h^2 \cos\phi_l \sin(2\phi_h)}, \quad (S3)$$

$$\frac{C_j}{A_j} = \frac{k_{l,j} n_h^2 \sin\phi_h \left[\cos(2\phi_l)\exp(i\phi_0)+1\right] + k_{h,j} n_l^2 \cos\phi_h \sin(2\phi_l)\exp(i\phi_0)}{k_{h,j} n_l^2 \sin\phi_l \left[\exp(i\phi_0)+\cos(2\phi_h)\right] + k_{l,j} n_h^2 \sin(2\phi_h)\cos\phi_l}, \quad (S4)$$

$$\frac{D_j}{A_j} = \frac{k_{l,j} n_h^2 \cos\phi_h \left[\cos(2\phi_l)\exp(i\phi_0)-1\right] - k_{h,j} n_l^2 \sin\phi_h \sin(2\phi_l)\exp(i\phi_0)}{k_{h,j} n_l^2 \sin\phi_l \left[\exp(i\phi_0)+\cos(2\phi_h)\right] + k_{l,j} n_h^2 \sin(2\phi_h)\cos\phi_l}, \quad (S5)$$

where $\phi_0 = k_{x0}\Lambda$, $\phi_l = k_{l,j} w_l$, and $\phi_h = k_{h,j} w_h$.

## 2. 2D Reflectance map for larger incident angle

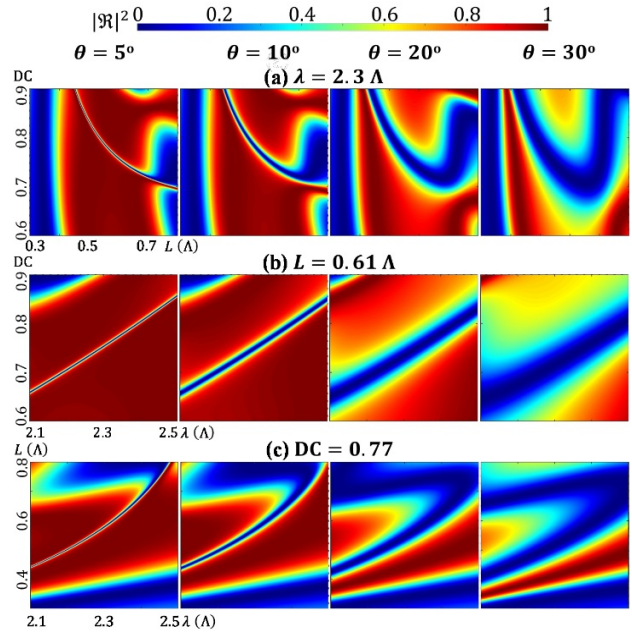

Fig. S1. Dispersion relations of the WAMs excited in the Si-air HCG at different incident angles $\theta$. Red, blue, and green are for TM$_0$, TM$_1$ and TM$_2$, respectively. Solid curves: 0°. Open circles: 5°. Dashes: 10°. Dots: 20°.